\documentclass[a4paper,11pt]{article}
\pdfoutput=1 

\usepackage{jinstpub} 

\title{\boldmath Calibration of a two-phase xenon time projection chamber with an \textsuperscript{37}Ar source}

\author[a,b,c,1]{E. M. Boulton,\note{Corresponding author}}
\author[a,b,c]{E. Bernard,}
\author[d,2]{N. Destefano,\note{Now at: The MITRE Corporation, 202 Burlington Rd, Bedford, MA 01730}}
\author[a,3]{B. N. V. Edwards,\note{Now at: IBM Research, STFC Daresbury Laboratory, Warrington, WA4 4AD, UK}}
\author[d]{M. Gai}
\author[a,b,c,4]{S. A. Hertel,\note{Now at: University of Massachusetts-Amherst, Dept. of Physics, 1126 Lederle Graduate Research Tower, Amherst, MA 01003-9337 USA}}
\author[a,5]{M. Horn,\note{Now at: Sanford Underground Research Facility, 630 E. Summit Street, Lead, SD 57754}}
\author[a,6]{N. A. Larsen,\note{Now at: University of Chicago, Kavli Institute for Cosmological Physics, 5640 Ellis Ave, Chicago, IL 60637}}
\author[a]{B. P. Tennyson,}
\author[a,7]{C. Wahl, \note{Now at: H3D, Inc., 3250 Plymouth Rd Suite 203, Ann Arbor, MI 48105}}
\author[a,b,c]{and D. N. McKinsey}

\affiliation[a]{Yale University, Dept. of Physics, 217 Prospect St., New Haven CT 06511, USA}
\affiliation[b]{University of California Berkeley, Department of Physics, Berkeley, CA 94720}
\affiliation[c]{Lawrence Berkeley National Laboratory, Berkeley, CA 94720}
\affiliation[d]{LNS at Avery Point, University of Connecticut, 1084 Shennecossett Rd. Groton, CT 06340}

\emailAdd{emboulton@lbl.gov}

\abstract{We calibrate a two-phase xenon detector at 0.27\,keV in the charge channel and at 2.8\,keV in both the light and charge channels using an \textsuperscript{37}Ar source that is directly released into the detector. We map the light and charge yields as a function of electric drift field. For the 2.8\,keV peak, we calculate the Thomas-Imel box parameter for recombination and determine its dependence on drift field. For the same peak, we achieve an energy resolution, $E_{\sigma}/E_{mean}$, between 9.8\% and 10.8\% for 0.1\,kV/cm to 2\,kV/cm electric drift fields. }


\arxivnumber{1705.08958} 

\begin{document}
\maketitle
\flushbottom

\section{Introduction}
\label{sec:Introduction}

\indent The two-phase xenon time projection chamber (TPC) has become a leading detector technology for weakly interacting massive particle (WIMP) \cite{Akerib:Results2016,Aprile:2012,Tan:2016}, axion, and axion-like particle (ALP) searches \cite{Aprile:2014,Akerib:Axions2017}. These detectors are also employed to detect coherent neutrino-nucleus scattering from nuclear reactors, spallation sources, and supernova neutrinos \cite{Santos:2011,Akimov,Horowitz:2003,Chakraborty:2014,Lang:2016}. Two-phase xenon TPCs are currently being explored for use in neutrinoless double beta decay, neutrino magnetic moment searches \cite{Baudis:2014,Coloma:2014}, and Compton imaging of gamma rays \cite{Wahl:2012}.\\ 
\indent The understanding of the charge and light signals in liquid xenon has developed substantially in the last few decades. The details of the mechanisms that generate the detectable signal are reviewed in \cite{henriqueReview,aprileReview}.\\
\indent When a particle interacts with xenon, it deposits its energy either as an electron recoil (ER) or a nuclear recoil (NR). The kinetic energy of the incoming particle results in some combination of excited xenon atoms, ionized xenon atoms, and heat. An excited atom pairs with a ground-state neighbor to form an excited dimer. An ionized atom can recombine with an electron, resulting similarly in an excited atom that forms a dimer. When the dimer de-excites, it gives off scintillation light with a wavelength of 178\,nm. This is the S1 signal, also referred to as the ``prompt" scintillation. A second scintillation signal, called S2 or ``proportional" scintillation, is created by the remaining electrons, which are drifted through the liquid to the surface by an electric field. They are extracted into the gaseous xenon by a stronger electric field. In the gas, the electrons interact with the gaseous atoms to create the second scintillation signal.  \\
\indent The fraction of xenon ions that recombine ($r$) affects both the number of photons produced ($n_{\gamma}=N_{ex}+N_{i}r$), which is related to the S1 signal, and the number of electrons produced ($n_{e}=N_{i}(1-r)$), which is related to the S2 signal. $N_{ex}$ and $N_{i}$ are the number of initially excited and ionized atoms. The recombination fraction can then be expressed as

\begin{equation} \label{eq1}
r = \frac{\frac{n_{\gamma}}{n_{e}}-\frac{N_{ex}}{N_{i}}}{\frac{n_{\gamma}}{n_{e}}+1}
\end{equation}

\indent Thomas and Imel model the rate at which the number of electrons and xenon ions change after the initial particle-xenon interaction with diffusion equations and include a term to represent recombination \cite{Thomas:1987}. Using properties of electron and xenon ion diffusion in xenon, they simplify these equations and apply a boundary condition that the electron-ion pairs are isolated and uniformly distributed in a box of dimension $a$. They derive 

\begin{equation} \label{eq2}
r = 1-\frac{1}{\xi}\ln(1+\xi)
\end{equation}

\begin{equation} \label{eq5}
\xi\equiv\frac{N_{i}\alpha}{4a^{2}\nu}
\end{equation}

\noindent to describe the recombination fraction, where $\alpha$ is a constant related to electron and ion mobility and $\nu$ is the mean ionized electron speed. \\
\indent Understanding xenon's response in the below-10\,keV regime is vitally important for quantifying the sensitivity of large scale dark matter detectors (such as LUX-ZEPLIN (LZ), XENON1T, and PandaX) to lower mass WIMPs \cite{Mount:2017,Aprile:2016,Tan:PRL2016}. Nuclear and electron recoils can be differentiated by their S2/S1 ratios, because ERs generally have a higher S2/S1 ratio than NRs for the same S1 area. Electron recoils that have a diminished S2/S1 ratio cannot be distinguished from nuclear recoils. Therefore, examining low-energy ERs will allow for a better understanding of discrimination at low energy between different types of recoils, and thus improve detector sensitivity to lower-energy WIMPs through background reduction. \\
\indent Recently, papers such as \cite{McCabe} and \cite{Kouvaris} have discussed the possibility of detecting sub-GeV WIMPs in two-phase xenon TPCs using the low-energy ER signal from the bremsstrahlung radiation produced from a WIMP interacting inelastically with a xenon nucleus. Additionally, understanding this low-energy regime will quantify the sensitivity of two-phase xenon detectors to axions, ALPs, and the neutrino magnetic moment. All of these are expected to produce low-energy ERs \cite{Akerib:Axions2017,Aprile:2014,Coloma:2014}.\\
\indent There is previous work studying the response of noble two-phase detectors to \textsuperscript{37}Ar, a source of mono-energetic, low-energy ER events. Sangiorgio et al. studied \textsuperscript{37}Ar in a two-phase argon detector \cite{Sangiorgio:2013}. They measured the charge yield of both the 0.27\,keV and 2.8\,keV peaks at a drift field of 2.4\,kV/cm, as well as the ratio of the two peaks. Akimov et al. studied several isotopes, including \textsuperscript{37}Ar, in a small, two-phase xenon detector \cite{Akimov:2014}. They measured the charge yield for the 2.8 keV peak at 3.75\,kV/cm. In this work, we present the effects of systematically varying electric drift field on the charge and light yield for \textsuperscript{37}Ar events in a two-phase xenon TPC. 

\section{\label{sec:PIXeY} PIXeY Detector}
\begin{figure}[h]
\begin{center}
\includegraphics[width=0.48\textwidth,clip]{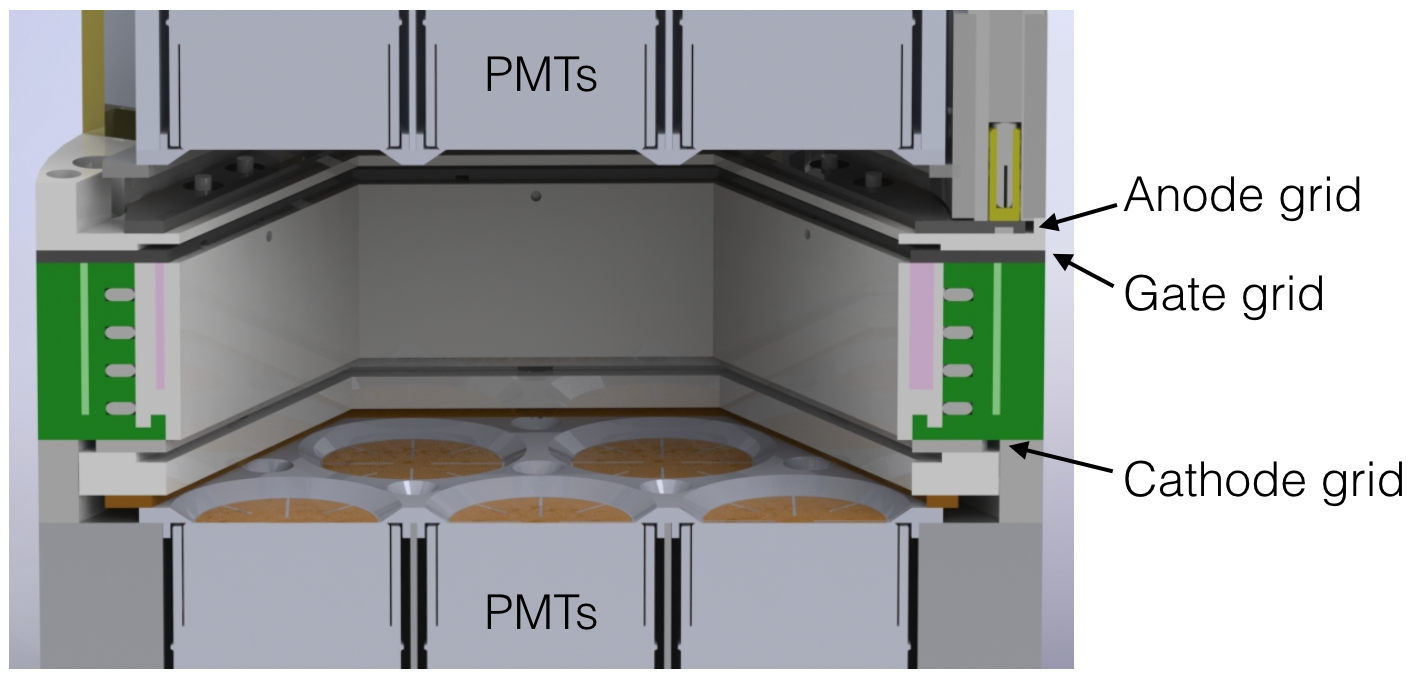}
\caption[]{\label{Schematic} Cutaway rendering of the PIXeY TPC. }
\end{center}
\end{figure}

PIXeY (Particle Identification in Xenon at Yale) is a hexagonal, two-phase xenon detector that was designed to measure the scintillation and ionization response of xenon as a function of drift field. Figure \ref{Schematic} shows a cross section of PIXeY. The potential difference between the cathode grid and the gate grid (5.05\,cm apart) forms the drift field and the potential difference between the gate grid and the anode grid (0.74\,cm apart) forms the extraction field. The detector is 18.4\,cm in width at its widest point. Although the detector holds approximately 12\,kg of xenon, only 2\,kg are included in the fiducial volume. The fiducial volume consists of a cylinder in the middle of the active liquid xenon. The diameter of this cylinder is 13\,cm, in order to avoid the edge effects of the drift field. A drift time cut of 3 to 30\,$\mu$s is applied in order to reduce backgrounds from the PMT array and electrodes. PIXeY was designed such that a strong electric field, up to 13.5\,kV/cm, can be applied to the extraction region in order to ensure nearly 100\% electron extraction efficiency. A wide range of voltages can be applied to the cathode, creating drift fields between 0.05 and 2\,kV/cm. In this experiment, the extraction field is 1.9\,kV/cm and the drift field is varied between 0.1 and 2\,kV/cm. In \cite{Edwards}, the electron extraction efficiency was measured to be $78\pm{5}$\,\%. The light collection efficiency, $g_{1}$, in PIXeY was calculated to be $9.7\pm{7}$\,\%, using a technique similar to \cite{Doke:2002}. However, instead of using several calibration sources with different energies at one drift field, one calibration source (\textsuperscript{83\it{m}}Kr) was used at several different drift fields. 

\indent The S1 and S2 signals are detected by two arrays of seven Hamamatsu R8778 photomultiplier tubes (PMTs) on the top and bottom of the TPC. The data acquisition (DAQ) system used a low-pass filter to trigger preferentially on small S2 pulses without triggering on every single photoelectron. The signal acquired from the PMTs undergoes an eight-fold amplification before it is digitized with a 12-bit ADC (CAEN V1720) at 250\,MHz. \\

\section{\label{sec:Source} \textsuperscript{37}Ar Source}
\textsuperscript{37}Ar decays via electron capture (EC) according to the decay reaction

\begin{equation}
\label{Ar37Prod}
^{37}\textrm{Ar}\rightarrow ^{37}\textrm{Cl} +\:\nu _{e}
\end{equation}

\noindent and releases x-rays from the capture of the K-shell, L-shell, and M-shell electrons at 2.8224\,keV, 0.2702\,keV, and 0.0175\,keV with branching ratios of 0.90, 0.09, and 0.009 \cite{Barsanov:2006}. These x-rays are photoabsorbed in the xenon bulk producing an electron recoil. \\
\indent \textsuperscript{37}Ar can be produced through reaction \ref{eqa} \cite{Michael:1984,Haxton:1988}, reaction \ref{eqb} \cite{Weber:1985,Kishore:1975}, or reaction \ref{eqc} \cite{SAGE:2006,CaO:2000,Barnes:1974,Barsanov:2006}. 

\begin{subequations}
\label{argonEqs}
\begin{align}
  \label{eqa}
  ^{36}\textrm{Ar}(\textrm{n,}\gamma)^{37}\textrm{Ar}\\  
  \label{eqb}
  ^{37}\textrm{Cl}(\textrm{p,n})^{37}\textrm{Ar} \\ 
  \label{eqc}
  ^{40}\textrm{Ca}(\textrm{n,}\alpha)^{37}\textrm{Ar}
\end{align}
\end{subequations}

\noindent In this work, reaction \ref{eqc} was used because it can be conveniently performed in a lab setting without a nuclear reactor, unlike reaction \ref{eqa}, or a high-energy proton beam, unlike reaction \ref{eqb}. \\
\indent An americium-beryllium (AmBe) source with $10^{6}$\,neutrons/s fluence was used to irradiate the calcium. The cross section of this reaction peaks at approximately 200\,mb for 5-6\,MeV neutrons \cite{Barnes:1974}. An AmBe source produces neutrons with energy up to 11 MeV \cite{Vega:2002}. It is possible to simply irradiate calcium powder, in the form of CaO, and \textsuperscript{37}Ar will evolve from the solid material; however, Abdurashitov et al., showed that argon does not begin to evolve until the powder is heated above 700\,$^{\circ}$C \cite{CaO:2000}. To avoid high-temperature baking, CaCl$_{2}$ was dissolved in water. Some of the \textsuperscript{37}Ar atoms produced in the solution through irradiation end up in the gas above the solution. Several purifying steps were necessary, because after irradiation the \textsuperscript{37}Ar is mixed with air and water vapor. The final \textsuperscript{37}Ar source needed to be free of oxygen and nitrogen to avoid degrading the PMT response when introduced to the TPC.

\begin{figure}[h]
\begin{center}
\includegraphics[width=0.48\textwidth,clip]{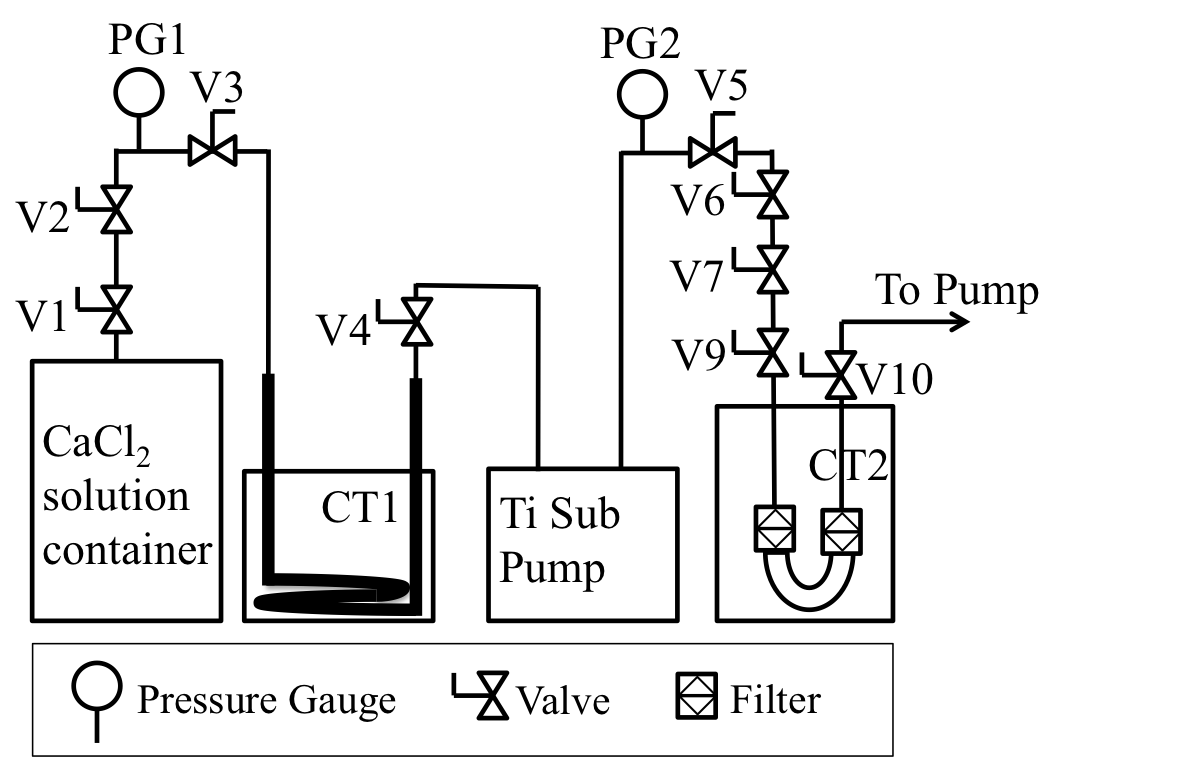}
\caption[]{\label{PurSys} Schematic of purification setup. V{\it x} refers to valves, PG{\it x} refers to pressure gauges, and CT{\it x} refers to cold traps.}
\end{center}
\end{figure}

\indent Before the CaCl$_{2}$ solution was irradiated, the solution was degassed through a series of three freeze-pump-thaw (FPT) cycles. During the pump step of each cycle, the pressure of the gas volume above the ice was reduced to $10^{-3}$\,Torr. The container was irradiated by the AmBe source for 21 days. Both the AmBe source and the container were inside a 7\,cm thick polyethylene cup. After irradiation, the solution container was attached to the purification system shown in figure \ref{PurSys}. The solution container and cold trap 1 (CT1) were placed in frozen ethanol baths (-116\,$^{\circ}$C) to freeze most of the water vapor. After several hours in equilibrium with CT1, the gas was introduced to the titanium sublimation pump, which consisted of a Gamma Vacuum PN360682 Titanium Filament kit inside a conflat container \cite{Gamma}. The Ti filament received repeated current pulses of 48 A, producing a layer of chemically active Ti on the inside walls of the conflat container. Non-noble impurities were efficiently removed by the Ti surface. Finally, the \textsuperscript{37}Ar was cryopumped into a charcoal-filled \textsuperscript{37}Ar storage region, cold trap 2 (CT2), which was submerged in a liquid nitrogen bath. CT2 was double valved on each side, removed from the purification setup, and attached into the PIXeY circulation system. In order to release the source into the detector, xenon gas was diverted through the charcoal trap before entering the detector, mixing \textsuperscript{37}Ar with the xenon flow. \\

\section{\label{sec:Results} Results}
\begin{figure}[h]
\begin{center}
\includegraphics[width=0.48\textwidth,clip]{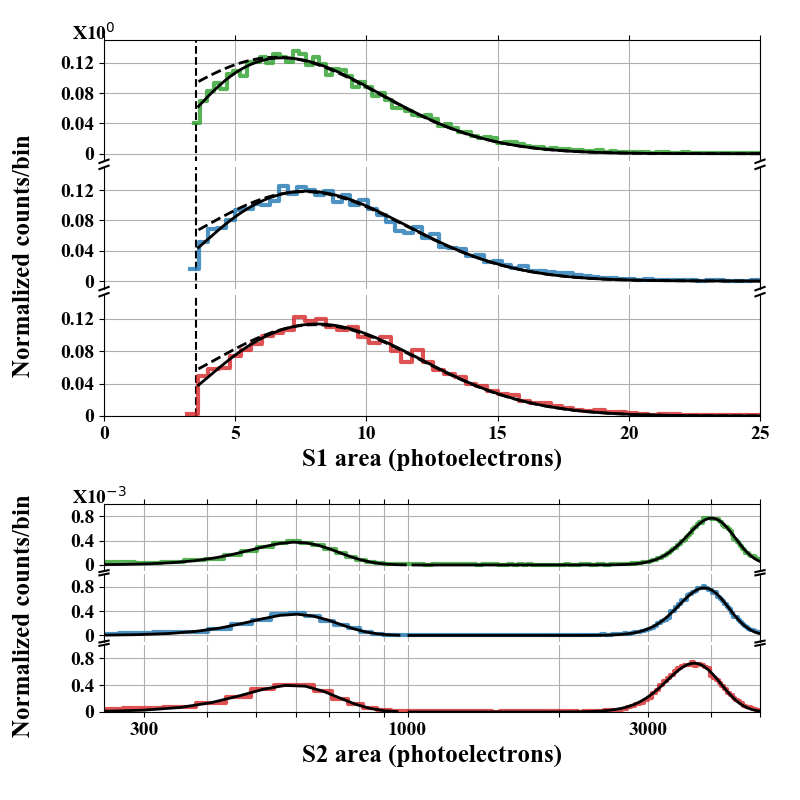}
\caption[]{\label{Ar_spectrum} [top] S1 area (phe) spectra at three fields. The peak is from the \textsuperscript{37}Ar K-shell x-ray at 2.8 keV. Fitted curves are as described in the text. [bottom] S2 area (phe) spectra at three fields. The left peak is from the L-shell x-ray at 0.27\,keV and the right peak is from the K-shell x-ray at 2.8\,keV \textsuperscript{37}Ar energy line. Each peak is fitted with a Gaussian function. For both top and bottom sets of spectra the red (bottom) spectra is at 0.099\,kV/cm, the blue (middle) is at 0.693\,kV/cm, and the green (top) is at 1.98\,kV/cm.}
\end{center}
\end{figure}

\indent Two Bq of \textsuperscript{37}Ar were introduced into the detector, and data were taken at six drift fields: 0.099, 0.198, 0.396, 0.693, 0.990, and 1.980\,kV/cm. For every data set, each recorded waveform was digitized and scanned by a pulse finding algorithm. Events containing zero or one S1 pulses and one S2 pulse were selected. The S1 and S2 areas in photoelectrons (phe) were determined by integrating the corresponding pulses, converting from ADC level-samples to electrons per photoelectron, and dividing by the average PMT single photoelectron response, $2.65\times10^{6}$ electrons per photoelectron. \\
\indent Several data quality selection cuts were applied to the events. A cut was applied to ensure that the S1 and S2 pulses contained the majority of the waveform's total pulse area. Another cut, based on pulse width, rejected events with large S1 pulses from gamma rays. \\
\indent The quantity $\log_{10}(\frac{\textrm{S2}}{\textrm{S1}})$ is used to discriminate between electron and nuclear recoils. The $\pm{1\sigma}$ curves that contain the majority of the ER events is referred to as the electron recoil band, and it is positioned above the nuclear recoil band in this discrimination space. In order to ensure that analyzed events were in the ER band, a selection on $\log_{10}(\frac{\textrm{S2}}{\textrm{S1}})$ was enforced. S1 and S2 spectra of the selected data are shown in figure \ref{Ar_spectrum}. 

\indent The S1 spectra exhibit a single peak, interpreted as the K-shell \textsuperscript{37}Ar peak at 2.8\,keV. Since this peak is on the order of single photoelectrons, one would expect the peak to be Poissonian. However, broadening of this peak was observed due to the non-uniform light collection and gain dispersion. Therefore a Poisson function was convolved with a Gaussian function and fit to the S1 spectra. For each $k$ number of events, the area of a Gaussian was set equal to the Poisson probability that those $k$ events occur in order to find the Gaussian amplitude. The convolved Poisson, a sum of Gaussians centered at $k$, is

\begin{equation} \label{eq3}
f(\textrm{S}1)=\sum_{k=1}^{\infty}\frac{\lambda^{k}e^{-\lambda}}{\sqrt{2\pi}k!\sigma}e^{-\frac{(\textrm{S}1-k)^{2}}{2\sigma^{2}}}
\end{equation}

\begin{figure}[h]
\begin{center}
\includegraphics[width=0.48\textwidth,clip]{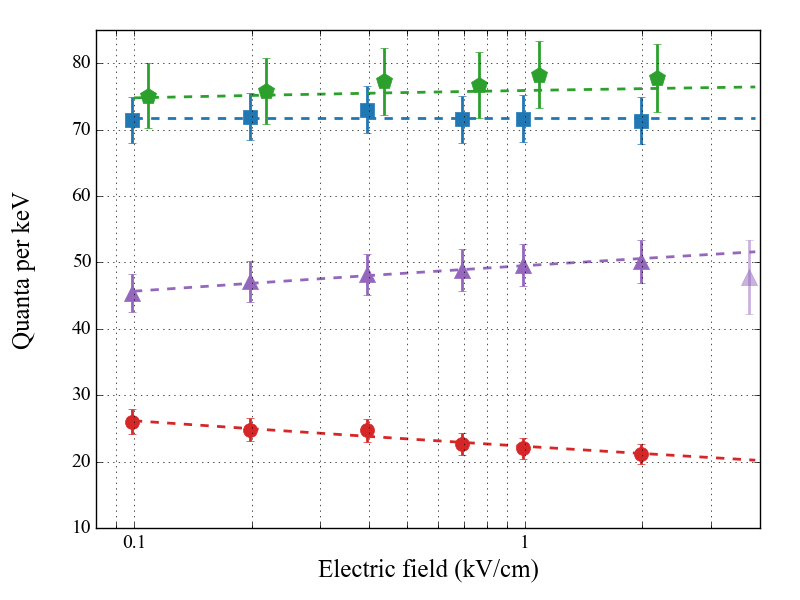}
\caption[]{\label{Yields} Yields at various fields for both \textsuperscript{37}Ar peaks. The red circles show the 2.8\,keV light yield. The purple triangles show the 2.8\,keV charge yield. The light purple triangle shows the 2.8\,keV charge yield measured by Akimov et al. \cite{Akimov:2014}. The blue squares show the total quanta for the 2.8\,keV peak. The green pentagons show the charge yield for the 0.27\,keV peak. The 0.27\,keV charge yield has been offset by 10\% in field for clarity. The dashed lines are calculated based on the Thomas-Imel box parameter versus field relationship.}
\end{center}
\end{figure}

\indent The DAQ trigger threshold prevented efficient measurement of small S1 pulses. For this reason, only the data between $\textrm{S}1=6$\,phe and $\textrm{S}1=15$\,phe were used to fit Eq.\,\ref{eq3}. To approximate the trigger efficiency, the value of the convolved Poisson function was divided by the data for each bin in the histogram and a logistic function was fit to the resulting points. The logistic function across all fields was similar: the S1 at which the logistic function reaches 50\% ranges from 2.97 to 3.20\,phe. The solid curves in figure \ref{Ar_spectrum} are the convolved Poisson function fitted for each field multiplied by the logistic threshold function averaged across all fields, while the dashed curves show just the convolved Poisson function. \\ 
\indent The light yield, measured in number of photons per keV, was computed from the S1 spectra. Because the convolved Poisson function was only fit on part of the peak, a one-dimensional $\chi^{2}$ minimization was performed on the convolved Poisson multiplied by the logistic threshold function by varying the Poisson parameter $\lambda$ for $\textrm{S}1=3$\,phe to $\textrm{S}1=15$\,phe. The measured light yield was the  $\lambda$ derived from this $\chi^{2}$ minimization divided by the two photoelectron correction, 1.175, the light collection efficiency, $g_{1}$, and 2.8\,keV. The two photoelectron correction was measured by \cite{Faham:2015} and assumed to be the same for PIXeY, since the same PMTs were used. The statistical uncertainty on the light yield was calculated from the $\chi^{2}$ minimization and the systematic uncertainty is dominated by the uncertainty in $g_{1}$. Figure \ref{Yields} shows the light yield for different drift fields.

\begin{figure}[h]
\begin{center}
\includegraphics[width=0.48\textwidth,clip]{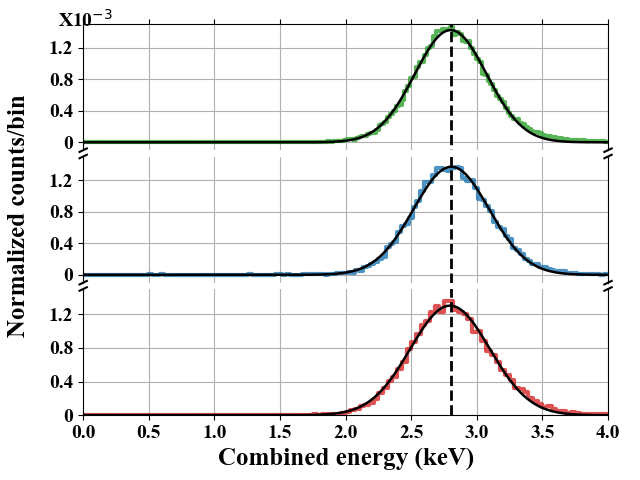}
\caption[]{\label{CombEnergy} Combined energy spectra for the 2.8 keV peak at three fields. Each peak is fitted with a Gaussian function. The red (bottom) spectra is at 0.099\,kV/cm, the blue (middle) is at 0.693\,kV/cm, and the green (top) is at 1.98\,kV/cm.}
\end{center}
\end{figure}

\indent The S1 area for the L-shell \textsuperscript{37}Ar peak at 0.27\,keV is expected to be 0.1-0.2\,phe from simulations. This is well below the DAQ trigger threshold, and therefore cannot be measured by PIXeY. \\
\indent The S2 spectra show the K-shell and the L-shell \textsuperscript{37}Ar peaks at 2.8\,keV and 0.27\,keV respectively. Because the charge signal is much larger than the light signal, these peaks were fit with a Gaussian function. The average branching ratio between the peaks across all fields is $0.128\pm{0.005}$, which agrees with the measurement by Sangiorgio et al., $0.116\pm{0.013}$ \cite{Sangiorgio:2013}, although it disagrees with earlier measurements, $0.103\pm{.003}$ and $0.098\pm{0.003}$ \cite{Santos:1960,Totzek:1967}. \\
\indent At each drift field, the charge yield, measured in number of electrons per keV, was calculated from the S2 spectrum; the mean of the Gaussian fit for each peak was divided by the single electron area, the electron extraction efficiency, and the peak energy. The single electron area was measured by fitting a Gaussian to the area histogram of small S2-like pulses, a population dominated by single electrons. The single electron area multiplied by the electron extraction efficiency is referred to as $g_{2}$, so that $\textrm{S2}=g_{2}n_{e}$. The statistical uncertainty on the charge yield is the error from the fitting. The systematic uncertainty is dominated by the uncertainty in the extraction efficiency. The charge yield as a function of the drift field for the 2.8 and 0.27\,keV peaks is shown in figure \ref{Yields}. \\
\indent A combined energy scale can be constructed from a linear combination of S1 and S2 so as to optimize the energy resolution, $E_{\sigma}/E_{mean}$. Figure \ref{CombEnergy} shows the K-shell, 2.8\,keV, peak in this combined energy scale, along with Gaussian fits. $E_{\sigma}/E_{mean}$ ranges from 9.8\% at 1.98\,kV/cm to 10.8\% at 0.099\,kV/cm. 

\begin{figure}[h]
\begin{center}
\includegraphics[width=0.48\textwidth,clip]{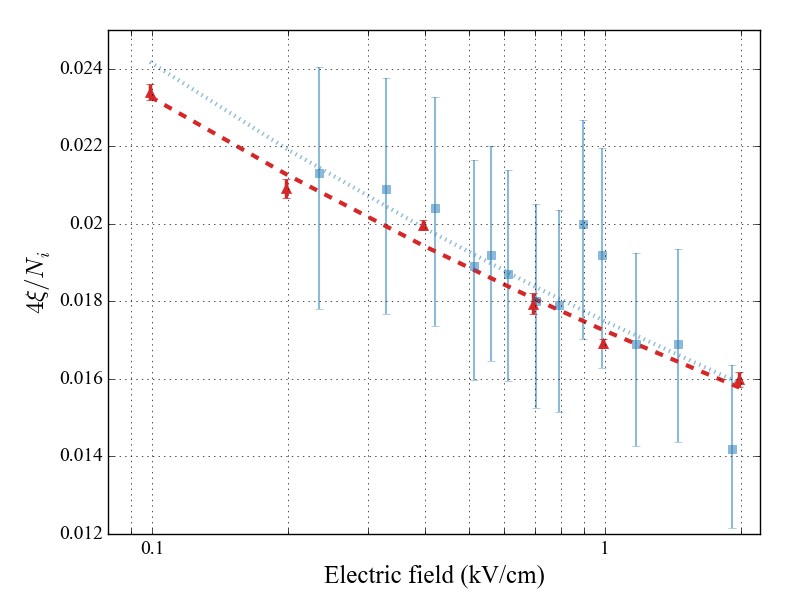}
\caption[]{\label{TIB} Thomas-Imel box ($4\xi/N_{i}$) parameter dependence on electric field. The red triangles are from this analysis. The blue squares are from Lin et al.\:analysis. Both lines are best fits to the data using Eq.\,\ref{eq6}.}
\end{center}
\end{figure}

\indent From Eq.\,\ref{eq1} it can be seen that the fraction of recombination can be calculated using the measured $n_{\gamma}$, $n_{e}$, and an estimate for $N_{ex}$/$N_{i}$. The quantity $N_{ex}$/$N_{i}$ cannot be measured directly, but it has been calculated by several analyses of two-phase xenon detectors \cite{Takahashi:1975,Doke:2002,Aprile:2007}. For this analysis, $N_{ex}$/$N_{i}$ was assumed to be 0.06 in order to match the work of Lin et al. as well as Noble Element Simulation Technique (NEST) \cite{Lin:2015,Szydagis:2013}. Once $r$ is calculated, the Thomas-Imel box parameter, $4\xi/N_{i}$, follows from Eqs.\,\ref{eq2} and \ref{eq5}. Figure \ref{TIB} shows the $4\xi/N_{i}$ dependence on field. The data are fit assuming the power-law dependence 

\begin{equation} \label{eq6}
\frac{4\xi}{N_{i}}=A\cdot\left(\frac{F*\textrm{cm}}{\textrm{kV}}\right)^{-\delta}
\end{equation}

\noindent where $F$ refers to electric drift field. Best-fit values are found to be $A=0.017\pm{0.002}$ and $\delta=0.130\pm{0.013}$. The dashed lines shown in figure \ref{Yields} are not fits but are expressions based on this fitted Thomas-Imel recombination model.

\section{\label{sec:Discussion} Discussion}
\begin{figure}[h]
\begin{center}
\includegraphics[width=0.48\textwidth,clip]{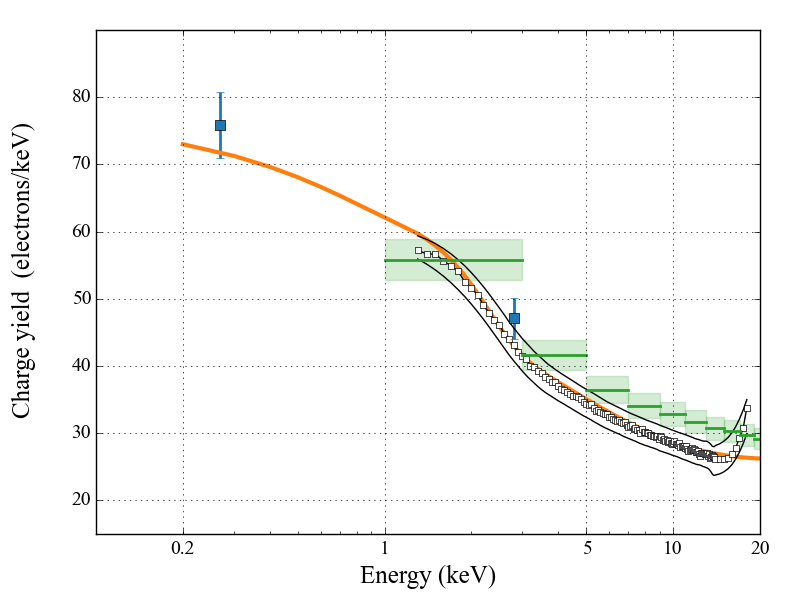}
\caption[]{\label{YieldvEnergy} Charge Yield versus Energy. The blue squares are the \textsuperscript{37}Ar points at 0.198\,kV/cm from this analysis. The green bands are charge yield measurements from Goetzke et al.\:at 0.19\,kV/cm. The band of black squares is from the tritiated methane measurements in LUX at .182\,kV/cm. The orange curve is the LZ TDR NEST package model for 0.18\,kV/cm \cite{Mount:2017}.}
\end{center}
\end{figure}

\indent We have characterized the ER response of a two-phase xenon detector using two mono-energetic low-energy peaks from \textsuperscript{37}Ar decay: the K-shell EC x-ray at 2.8\,keV and the L-shell EC x-ray at 0.27\,keV.  The detector was able to identify the S1 and S2 signals from the K-shell x-ray and S2 signal from the L-shell x-ray for six electric drift fields.  \\
\indent The charge yields for the 0.27 and 2.8\,keV peaks at 0.198\,kV/cm agree with the charge yields for other low-energy points, shown in figure \ref{YieldvEnergy}. In LUX, tritiated methane was injected and a band of charge yield from 1.3 to 18\,keV was measured at 0.182\,kV/cm \cite{Akerib:tritium2016}. Goetzke et al. measured the charge yield for a range of energies from a 661.7\,keV gamma $^{137}$Cs source using the Compton coincidence technique at 0.19\,kV/cm, also shown in figure \ref{YieldvEnergy} \cite{Goetzke}. \\ 
\indent NEST has been developed using a collection of models to explain the light and charge yields of noble elements \cite{Szydagis:2013}. It assumes a constant work function or $W$ value, the energy required to produce a photon or electron, of 13.7\,eV and the Thomas-Imel box model for recombination at low energies. This paper's analysis is consistent with NEST's prediction. 

\indent Additionally, if the $W$ value is constant down to 0.2 keV as NEST assumes, one would expect the total quantum yield (sum of the charge yield and light yield) for the 2.8 and 0.27\,keV peaks to be the same. This analysis was only able to calculate the total quantum yield for the K-shell peak, which varied between 71.3$\pm{3.2}$ and 73.0$\pm{3.1}$; however, with $N_{ex}/N_{i}=0.06$, the recombination model predicts a total quantum yield for the L-shell peak between 79.6$\pm{4.9}$ and 83.0$\pm{5.1}$ for the various drift fields. Comparing the K-shell peak total quantum yield to the L-shell peak estimated quantum yield, the total quantum yields are statistically consistent across all drift fields, given the uncertainties in $g_{1}$ and $g_{2}$. This is consistent with NEST's assumption of a constant $W$ value.\\   
\indent The energy resolutions calculated from these PIXeY data improve upon the energy resolution predictions for 2.8\,keV from previous experiments. In analyses of LUX, XENON10, and MiX data the function $E_{\sigma}/E_{mean}\cdot 100\%=a/\sqrt{E}+b$ was fit to energy resolutions for different energy peaks \cite{Akerib:ER2017,Aprile:2011,Stephenson:2015}. In the LUX analysis, $a=33\pm{0.01}$\,keV$^{-1/2}$ and $b=0$, in the XENON10 analysis, $a=44.3$\,keV$^{-1/2}$ and $b=0$, in the MiX analysis, $a=22$\,keV$^{-1/2}$ and $b=0.42$. Using these fits for 2.8 keV, energy resolutions of 19.7\%, 26.5\%, and 13.7\% are expected from the LUX, XENON10, and MiX analyses, compared to the values of 9.8\% to 10.8\% measured here.\\
\indent We also investigated the relationship of recombination with drift field. The LUX tritium data show that recombination decreases with energy \cite{Akerib:tritium2016}. This PIXeY \textsuperscript{37}Ar analysis shows results consistent with that finding. For the 2.8\,keV peak, as the field increases the charge yield increases, indicating that recombination decreases as the field increases. However, for the 0.27\,keV peak, as the field increases the charge yield stays nearly constant. This indicates that recombination is small and does not vary substantially with field at 0.27\,keV. For the K-shell peak, $r=0.288\pm{0.025}$ and for the L-shell peak, $r=0.023\pm{0.021}$, which is consistent with zero.\\
\indent The calculation of the Thomas-Imel box parameter agrees with an analysis from Lin et al. The Lin analysis calculates $4\xi/N_{i}$ by comparing simulations to data and then fits the relationship between $4\xi/N_{i}$ and field to Eq.\,\ref{eq6} \cite{Lin:2015}. Lin et al.'s analysis found $A=0.017\substack{+0.003 \\ -0.002}$ and $\delta=0.140$, which are similar to the values found in this analysis, $A=0.017\pm{0.002}$ and $\delta=0.130\pm{0.013}$. Additionally, the Akimov et al. 2.8\,keV charge yield measurement of $47\pm{5.5}$\,electrons/keV at 3.75\,kV/cm matches within error bars the measured charge yield field dependence shown in figure \ref{Yields} \cite{Akimov:2014}.\\

\acknowledgments
This work is supported by DHS grant 2011-DN-007-ARI056, NSF grants CBET-1039206 and PHY-1312561, and DOE grant DE-FG02-94ER40870. This material is also based upon work supported by the U.S. Department of Energy, Office of Science, Office of Workforce Development for Teachers and Scientists, Office of Science Graduate Student Research (SCGSR) program. The SCGSR program is administered by the Oak Ridge Institute for Science and Education (ORISE) for the DOE. ORISE is managed by ORAU under contract number DE‐AC05‐06OR23100. This work represents the views and opinions of the authors and not those of the MITRE corporation.

\paragraph{Note added.} This is also a good position for notes added
after the paper has been written.


\end{document}